\documentclass[prb,twocolumn]{revtex4}
\usepackage{graphicx}

\begin{document}

\title[]{Multistability and Self-Organization in Disordered SQUID Metamaterials} 

\author{N. Lazarides, G. P. Tsironis}
\affiliation{
Department of Physics, University of Crete, P.O. Box 2208,
71003 Heraklion, Greece, $\&$
Institute of Electronic Structure and Laser,
Foundation for Research and Technology-Hellas, P.O. Box 1527,
71110 Heraklion, Greece \\
nl@physics.uoc.gr, gts@physics.uoc.gr}
\begin{abstract}
Planar arrays of magnetoinductively coupled rf SQUIDs (Superconducting Quantum 
Interference Devices) belong to the emergent class of superconducting metamaterials 
that encompass the Josephson effect. These SQUID-based metamaterials acquire their  
electromagnetic properties from the resonant characteristics of their constitutive 
elements, i.e., the individual rf SQUIDs. In its simplest version, an rf SQUID 
consists of a superconducting ring interrupted by a Josephson junction.
We investigate the response of a two-dimensional rf SQUID metamaterial to frequency 
variation of an externally applied alternating magnetic field in the presence of 
disorder arising from critical current fluctuations of the Josephson elements; 
in effect, the resonance frequencies of individual SQUIDs are distributed randomly 
around a mean value. Bistability is observed in the total current-frequency curves 
both in ordered and disordered SQUID metamaterials; moreover, bistability is favoured
by disorder through the improvement of synchronization between SQUID oscillators.
Relatively weak disorder widens significantly the bistability region by helping the
system to self-organize itself and leads to
nearly homogeneous states that change smoothly with varying driving frequency.
Also, the total current of the metamaterial is enhanced compared with that of 
uncoupled SQUIDs, through the synergetic action of coupling and synchronization.
The existence of simultaneously stable states that provide either high or low 
total current, allows the metamaterial to exhibit different magnetic responses
that correspond to different values of the magnetic permeability. 
and provide either high or low total current, allows the metamaterial to exhibit 
two different magnetic responses that correspond to different values of its magnetic
permeability. At low power of the incident field, high-current states exhibit
extreme diamagnetic properties corresponding to negative magnetic permeability
in a narrow frequency region.
\end{abstract}

\maketitle


\section{Introduction}
Advances in theory and nanofabrication techniques have opened many opportunities 
for researchers to create artificialy structured, composite media that exhibit
extraordinary properties. The {\em metamaterials} (MMs) are perhaps the most 
representative class of materials of this type,
which, among other fascinating properties, exhibit negative refractive index and 
optical magnetism \cite{Shalaev2007,Soukoulis2007,Litchinitser2008}.
High-frequency magnetism, in particular, exhibited by the {\em magnetic metamaterials},
is considered one of the 'forbidden fruits' 
in the Tree of Knowledge that has been brought forth by metamaterial research
\cite{Zheludev2010}.
The unique properties of MMs are particularly well suited for novel devices like 
hyperlenses, which surpass the diffraction limit \cite{Pendry2000}, and optical 
cloaks of invisibility \cite{Schurig2006}. Furthermore, they can form a material
base for other functional devices with tuning and switching capabilities
\cite{Zheludev2010,Zheludev2011}. The key element for the construction of MMs
has customarily been the split-ring resonator (SRR), a subwavelength "particle"
which is effectivelly a kind of an artificial "magnetic atom" \cite{Caputo2012}. 
In its simplest version it is just a highly conducting ring with a slit that can
be regarded as an inductive-capacitive resonant oscillator. SRRs become nonlinear
and therefore tunable with the insertion of an electronic component (e.g., a diode) 
in their slits \cite{Shadrivov2008}. 
However, metallic SRRs suffer from high ohmic losses
that place a strict limit on the performance of SRR-based metamaterials, either 
in the linear or the nonlinear regime, and hamper their use in novel devices.
The incorporation of active constituents in metamaterials that provide gain
through external energy sourses has been recognized as a promising technique
for compensating losses \cite{Boardman2010}. On the other hand, the replacement of
the metalic elements with superconducting ones, provides both loss reduction 
and wideband tuneability \cite{Anlage2011}; the latter because of the extreme 
sensitivity of the superconducting state to external stimuli 
\cite{Zheludev2011,Anlage2011}. 
Tunability of superconducting metamaterial properties by varying the temperature 
or an externally applied magnetic field have been recently demonstrated
\cite{Ricci2005,Ricci2007,Gu2010,Fedotov2010,Chen2010}. 

Going a step beyond, the metalic metamaterial elements can be replaced by rf SQUIDs 
(rf Superconducting QUantum Interference Devices), creating thus SQUID-based
metamaterials \cite{Lazarides2007,Lazarides2008}. The rf SQUID, as shown in figure 1(a), 
consists of a superconducting ring interrupted by a Josephson junction (JJ) and 
demonstrates both reduced losses and strong nonlinearities due to the Josephson 
element \cite{Barone1982,Likharev1986}. It constitutes the direct superconducting 
analogue of a nonlinear SRR, that plays the role of the 'magnetic atom' in 
superconducting metamaterials in a way similar to that of the SRR for conventional
(metalic) metamaterials \cite{Lazarides2007,Tsironis2009,Lazarides2012}. 
However, currents and voltages in SQUID elements are determined by the celebrated
Josephson relations \cite{Josephson1962}. 
The feasibility of using superconducting circuits with 
Josephson junctions as basic elements for the construction of superconducting 
thin-film metamaterials has been recently demonstrated \cite{Jung2013}.
Nonlinearity and discreteness in SQUID-based metamaterials may also lead in the 
generation of nonlinear excitations in the form of discrete breathers  
\cite{Lazarides2008,Tsironis2009,Lazarides2012}, time-periodic and spatially 
localized modes that change locally their magnetic response.
Numerous types of SQUIDs have been investigated since its discovery, while it has
found several technological applications \cite{Kleiner2004,Fagaly2006}. 
Recent advances that led to nano-SQUIDs makes possible
the fabrication of SQUID metamaterials at the nanoscale \cite{Wernsdorfer2009}. 
The use of SQUID arrays 
in dc current sensors \cite{Beyer2008}, filters \cite{Haussler2001a,Bruno2004},
magnetometers \cite{Matsuda2005}, amplifiers \cite{Hirayama1999,Huber2001,Castellanos2007},
radiation detectors \cite{Kaplunenko1993}, flux-to-voltage converters \cite{Haussler2001b}, 
as well as in rapid single flux quantum (RFSQ) electronics \cite{Brandel2012},
has been suggested and realized in the past. However, in most of these works the SQUIDs
in the arrays were actually directly coupled through conducting paths. 
The SQUID-based metamaterial suggested in reference \cite{Lazarides2007} and 
investigated further in the present work relies on 
the magnetic coupling of its elements through dipole-dipole forces due to the mutual
inductance between SQUIDs.
\begin{figure}[!h]
\includegraphics[angle=0, width=0.80\linewidth]{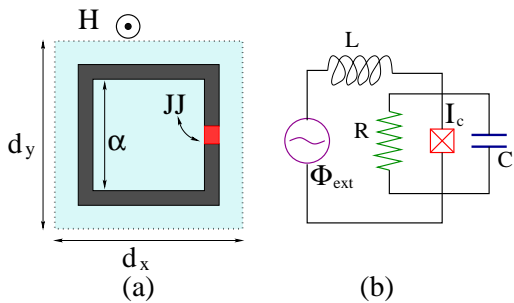}
\caption{(color online)
(a) Schematic drawing of an rf SQUID in a perpendicular time-dependent magnetic field
${\bf H}(t)$.  
(b) Equivalent electrical circuit ($RCSJ$ model) for a single rf SQUID driven by a 
flux source $\Phi_{ext}$.
}
\end{figure}

Moreover, at low (sub-Kelvin) temperatures provide access to the quantum regime, 
where rf SQUIDs can be manipulated as flux and phase qubits
\cite{Poletto2009,Castellano2010}, 
the basis element for quantum computation. Inductively coupled SQUID flux qubits
can be used for the realization of quantum gates \cite{Zhou2005}, while larger 
arrays of inductively coupled SQUID flux qubits have been proposed as scalable 
systems for adiabatic quantum computing 
\cite{Roscilde2005,Johnson2010}.   
Notably, amplification and squeezing of quantum noise has been recently achieved 
with a tunable SQUID-based metamaterial \cite{Castellanos2008}.

In the present work we investigate the response of a two-dimensional (2D) rf SQUID 
metamaterial in the planar geometry with respect to frequency variation of an 
alternating magnetic field, focusing on the effect of quenched disorder through the 
SQUID parameter $\beta$ that determines the resonance frequency of individual
SQUIDs. We are particularly interested in frequencies near resonance where multiple
responses may appear; 
this issue is related to the existence or not of a bistability region in the
current-frequency curve of the SQUID metamaterial. Having calculated the response 
of the metamaterial to a given alternating filed, the effective magnetic permeability 
$\mu_r$ can be determined by temporal and spatial averaging.
Different responses, corresponding to different simultaneously stable metamaterial
states, result in different $\mu_r$ in the bistability region. 
In the next section we give a brief overview of the equations for a single 
SQUID, we introduce the equations for the flux dynamics in a 2D SQUID metamaterial
and calculate its linear modes. In Section 3 we present numerical results for the
maximum total current (devided by the total number of SQUIDs) as a function of the 
driving frequency. Such current-frequency curves are obtained both for ordered and 
disordered SQUID metamaterials, focusing primarily on the possibility of bistability.
In Section 4 we discuss the effect of synchronization and present calculations of 
the relative magnetic permeability $\mu_r$. Section 5 contains the conclusions.

\section{Dynamic Equations and Linear Modes
}
In an ideal JJ, the current-phase relation is of the form $I=I_c \, \sin(\phi_j)$,
where $I_c$ is the critical current of the JJ and $\phi_j$ the Josephson phase. 
When driven by an external magnetic field $H(t)$, the induced (super)currents 
around the SQUID ring are determined by the celebrated Josephson relations
\cite{Josephson1962}.
In the equivalent circuit picture, the resistively and capacitively shunted 
junction (RCSJ) model is frequently adopted to describe a real JJ. 
Thus, the equivalent lumped circuit for the rf SQUID in a magnetic field with
appropriate polarization comprises a flux source $\Phi_{ext}$
in series with an inductance $L$ and an ideal JJ, while the latter is shunted
by a capacitor $C$ and a resistor $R$ [figure 1(b)]. 
Then, the dynamic equation for the flux threading the SQUID ring can be obtained
by application of Kirkhhoff laws, as
\begin{equation}
\label{2.01}
 C \frac{d^2 \Phi}{dt^2} +\frac{1}{R} \frac{d \Phi}{dt} 
                         +I_c \, \sin\left( 2 \pi \frac{\Phi}{\Phi_0} \right)
                         + \frac {\Phi -\Phi_{ext}}{L} =0, 
\end{equation}	
where $\Phi_{ext}$ is the external flux, $\Phi_0$ is the magnetic flux quantum, 
and $t$ is the temporal variable.
The flux $\Phi$ threading the SQUID ring is related to the Josephson phase through
the flux quantization condition
\begin{equation}
\label{2.02}
  \phi_j =2 \pi \frac{\Phi}{\Phi_0} +2 \pi n,
\end{equation}
where $n$ can be any integer.
\begin{figure}[!t]
\includegraphics[angle=-0, width=0.80\linewidth]{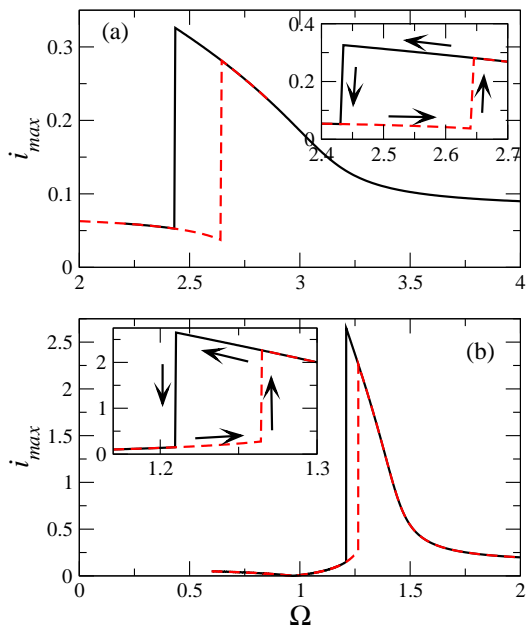}
\caption{(color online)
Maximum current amplitude $i_{max}$ as a function of the driving frequency
$\Omega$ for an rf SQUID with $\alpha=0.002$, $\phi_{dc} =0$, and 
(a) $\beta=1.27$, $\phi_{ac} =0.1$; (b) $\beta=0.15$, $\phi_{ac} =0.02$. 
An enlargement of the bistability region is shown in the insets for each case.
}
\end{figure}

The rf SQUID is a highly nonlinear oscillator whose amplitude-frequency
curves exhibit several peculiar features not seen in conventional inductive-capacitive
oscillators. For example, in a fine scale one may observe internal structure
of increasing complexity that increases with increasing $\beta$ \cite{Shnyrkov1980}. 
That fine structure of the amplitude-frequency curves can be reproduced numerically by
integrating the dynamic equation (\ref{2.01}) \cite{Lazarides2008,Lazarides2012}.
The SQUID resonance can be tuned either by varying the amplitude of the alternating
driving field or by varying the magnitude of a static (dc) field threading the SQUID
ring that creates a flux bias. 
The resonance shift due to nonlinearity has been actually observed in a Josephson
parametric amplifier driven by fields of different power levels \cite{Castellanos2007},
while the shift with applied DC flux has been seen in high$-T_c$ rf SQUIDs \cite{Zeng2000}
and very recently in a low$-T_c$ rf SQUID in the linear regime \cite{Jung2013}. 
Systematic measurements on microwave resonators comprising SQUID arrays
are presented in references \cite{Castellanos2007,Palacios2008}.
For very low amplitude of the driving field (linear regime), the rf SQUID exhibits 
a resonant magnetic response at a particular frequency 
$\omega_{SQ} = \omega_0 \sqrt{ 1 +\beta_L }$, where $\omega_0 =1 / \sqrt{L C}$ 
is the inductive-capacitive SQUID frequency, and $\beta_L$ is the SQUID parameter
\begin{equation}
\label{2.03}
  \beta_L= 2\pi \beta =2\pi \frac{L I_c}{\Phi_0} .
\end{equation}

The dynamic behavior of the rf SQUID has been studied extensively for more than 
two decades both in the hysteretic ($\beta_L >1$) and the non-hysteretic
regimes, usually under an external flux field of the form
\begin{eqnarray}
  \label{2.05}
     \Phi_{ext} = \Phi_{dc}  +\Phi_{ac} \cos(\omega t ) ,
\end{eqnarray}
where $\omega$ is the driving frequency. The first and second term on the 
right-hand-side of the earlier equation correspond to the fluxes due to the
presence of a constant (dc) and an alternating (ac) spatially uniform magnetic 
field, respectively.
Typical current amplitude-frequency curves for a single SQUID are shown in 
figure 2 (for $\Phi_{dc} =0$).
Equation (\ref{2.01}) is formally equivalent to that of a massive particle 
in a tilted washboard potential (figure 3) 
\begin{eqnarray}
\label{2.06}
  U_{SQ} = \frac{1}{C} \left\{ \frac{(\Phi - \Phi_{ext})^2}{2 L} 
          -E_j \, \cos\left(2\pi \frac{\Phi}{\Phi_0}\right) \right\} ,
\end{eqnarray}
with $E_j = I_c \Phi_0 / (2\pi)$ being the Josephson energy. 
While for $\beta_L < 1$ there the potential has a single minimum, 
it aquires more and more local minima as $\beta_L$ increases above unity. 
Moreover, applied dc flux moves the location of both the local and the global 
minima.    
\begin{figure}[!h]
\includegraphics[angle=0, width=0.80\linewidth]{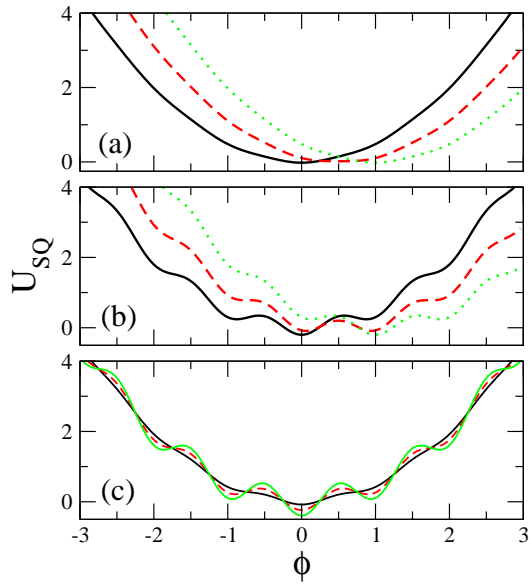}
\caption{(color online)
Potential curves as a function of the flux threading the SQUID ring.
(a) For a non-hysteretic SQUID with $\beta_L \simeq 0.75 <1$ and 
$\phi_{dc}=\Phi_{dc} /\Phi_0 =0$ (black-solid curve);
$0.5$ (red-dashed curve); $1.0$ (green-dotted curve). 
(b) For a hysteretic SQUID with $\beta_L \simeq 8 > 1$ and
$\phi_{dc}=\Phi_{dc} /\Phi_0 =0$ (black-solid curve);
$0.5$ (red-dashed curve); $1.0$ (green-dotted curve). 
(c) For $\Phi_{dc} =0$ and $\beta_L =0.5 <1$ (black-solid curve);
$1.5$ (red-dashed curve); $2.5$ (green-dotted curve).
}
\end{figure}

Consider a planar array comprising identical rf SQUIDs arranged in a tetragonal 
$N_x \times N_y$ lattice, that is placed in a spatially uniform, alternating
magnetic field directed perpendicularly to the plane of the SQUIDs.
The flux threading each SQUID ring induces a current with both normal and 
superconducting components that generates its own magnetic field. 
The induced fields couple the SQUIDs to each other through magnetic dipole-dipole 
interactions.  The strength of this magnetoinductive coupling
falls off approximatelly as the inverse-cube of the distance between SQUIDs.
If the distance between neighboring SQUIDs is such that they are weakly coupled,
then next-nearest and more distant neighbor coupling can be neglected.
In that case, we need to take into account nearest-neighbor interaction only 
between SQUIDs, and the flux threading $(n,m)-$th SQUID of the array is given by 
\begin{eqnarray}
\label{2.11}
  \Phi_{n,m} =\Phi_{ext} + L \, \left[ I_{n,m} +\lambda_x ( I_{n-1,m} + I_{n+1,m} ) \right.
     \nonumber \\
  \left.   +\lambda_y ( I_{n,m-1} + I_{n,m+1} ) \right] ,
\end{eqnarray}
where $n=1,...,N_x,\, m=1,...,N_y$, $I_{n,m}$ is the total current induced in the 
$(n,m)-$th SQUID, and 
$\lambda_{x,y} \equiv M_{x,y} / L$ are the magnetic coupling constants between
neighboring SQUIDs in the $x$ and $y$ directions, respectively.
The values of the $M_x$ and $M_y$ are negative since the magnetic field generated
by the induced current in a SQUID crosses the neighboring SQUID in the opposite 
direction. By adopting the resistively and capacitively shunted junction (RCSJ)
model, the current $I_{n,m}$ is
\begin{equation}
\label{2.12}
  -I_{n,m} = C\frac{d^2 \Phi_{n,m}}{dt^2} +\frac{1}{R} \frac{d \Phi_{n,m}}{dt} 
    + I_c\, \sin\left( 2\pi\frac{\Phi_{n,m}}{\Phi_0} \right) .
\end{equation}
Then, following the procedure of reference \cite{Lazarides2008} and neglecting terms
of order $\lambda_x \lambda_y$, $\lambda_y^2$, $\lambda_x^2$, etc., we get 
\begin{eqnarray}
  \label{2.13}
    C\frac{d^2 \Phi_{n,m}}{dt^2} +\frac{1}{R} \frac{d \Phi_{n,m}}{dt}
    + I_c\, \sin\left( 2\pi\frac{\Phi_{n,m}}{\Phi_0} \right)
    \nonumber \\
    -\lambda_x ( \Phi_{n-1,m} + \Phi_{n+1,m} )
    -\lambda_y ( \Phi_{n,m-1} + \Phi_{n,m+1} )
    \nonumber \\
    =[1 -2 (\lambda_x +\lambda_y )] \Phi_{ext} .
\end{eqnarray}    
In the absence of losses ($\gamma=0$), the earlier equations can be obtained from 
the Hamiltonian function
\begin{eqnarray}
\label{2.14}
   H = \sum_{n,m} \frac{Q_{n,m}^2}{2 C}
    \nonumber \\
    +\sum_{n,m} 
     \left[\frac{1}{2L} (\Phi_{n,m} -\Phi_{ext} )^2
      -E_J \, \cos\left( 2\pi \frac{\Phi_{n,m}}{\Phi_0} \right) \right]
  \nonumber \\
   -\sum_{n,m} \frac{\lambda_x}{L} (\Phi_{n,m} -\Phi_{ext} ) (\Phi_{n-1,m} -\Phi_{ext} )
  \nonumber \\
   -\sum_{n,m} \frac{\lambda_y}{L} (\Phi_{n,m} -\Phi_{ext} ) (\Phi_{n,m-1} -\Phi_{ext} ) , 
\end{eqnarray}
and 
\begin{eqnarray}
  \label{2.15}
    Q_{n,m} =C\, \frac{d \Phi_{n,m}}{dt}
\end{eqnarray}
\begin{figure}[!t]
\center{
\includegraphics[angle=-0, width=0.80\linewidth]{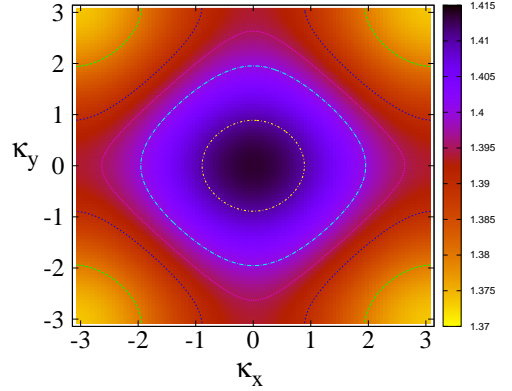}
}
\caption{(color online)
Contours of the linear dispersion $\Omega_{\vec\kappa}$ 
on the $\kappa_x - \kappa_y$ plane for a two-dimensional rf SQUID array,
with  $\lambda_x = \lambda_y =-0.014$ and $\beta=0.15$.
}
\end{figure}
is the canonical variable conjugate to $\Phi_{n,m}$, and represents the charge
accumulating across the capacitance of the JJ of each rf SQUID.
The Hamiltonian function (\ref{2.14}) is the weak coupling version of that 
proposed in the context of quantum computation \cite{Roscilde2005}.
Using the relations 
\begin{eqnarray}
\label{2.16}
  \phi_{n,m} = \frac{\Phi_{n,m}}{\Phi_0}, \phi_{ext} = \frac{\Phi_{ext}}{\Phi_0},
  \tau=\omega_0 t , \Omega=\frac{\omega}{\omega_0} ,  
\end{eqnarray}    
and equations (\ref{2.03}), (\ref{2.04}), equations (\ref{2.13}) are normalized to 
\begin{eqnarray}
\label{2.17}
  \ddot{\phi}_{n,m} +\gamma \dot{\phi}_{n,m} +\phi_{n,m}
   +\beta\, \sin( 2 \pi \phi_{n,m} )
 \nonumber \\
  -\lambda_x ( \phi_{n-1,m} +\phi_{n+1,m} ) 
  -\lambda_y ( \phi_{n,m-1} +\phi_{n,m+1} )
 \nonumber \\
   = \phi_{eff} , 
\end{eqnarray}	
where the overdots denote differentiation with respect to the normalized time $\tau$,
\begin{eqnarray}
\label{2.18}
  \phi_{eff} =[1-2(\lambda_x +\lambda_y)] \phi_{ext} ,   
\end{eqnarray}   
is the effective driving field, and 
\begin{equation}
\label{2.04}
  \gamma=\frac{1}{R} \sqrt{ \frac{L}{C} } ,
\end{equation}
is the loss coefficient of individual SQUIDs, that actually represents all of the
dissipation coupled to each rf SQUID and may also include radiative losses 
\cite{Kourakis2007}.

SQUID metamaterials support magnetoinductive flux-waves \cite{Lazarides2008},
just like conventional metamaterials comprising metallic elements 
(i.e., split-ring resonators) \cite{Shamonina2002}.
The frequency dispersion for small amplitude flux waves is obtained by  
the substitution of  
$\phi= A\, \exp[i (\kappa_x n + \kappa_y m - \Omega \tau)]$,
into the linearized equation (\ref{2.16}) without losses and external field 
($\gamma=0$, $\phi_{ext} =0$)
\begin{eqnarray}
  \label{2.21}
   \Omega = \sqrt{1 + \beta_L -2( \lambda_x \, \cos \kappa_x
                                +\lambda_y \, \cos \kappa_y ) }  , 
\end{eqnarray}
where $\Omega = \omega / \omega_0$ and $\kappa_{x,y} = d_{x,y} \, k_{x,y}$
are the normalized wavevector components, with $k_{x}$ ($k_{y}$) and $d_{x}$
($d_{y}$) being the wavevector component and center-to-center distance between
neighboring SQUIDs in $x-$direction ($y-$direction), respectively.

\section{Current - Frequency Curves}
In the following, equations (\ref{2.17}) are implemented with the boundary
condition (except otherwise stated)
\begin{eqnarray}
\label{2.22}
  \phi_{0,m} (\tau) =\phi_{N_x+1,m} (\tau) =\phi_{n,0} (\tau) =\phi_{n,N_y+1} (\tau)= 0 ,
\end{eqnarray}
for $n=1,...,N_x$, $m=1,...,N_y$ that account for the termination
of the structure in a finite system. The set of dynamic equations (\ref{2.22})
are integrated in time with a standard 4th order Runge-Kutta algorithm.
For a single SQUID, the current denoted by $i_{max}$ is just the amplitude of
the induced current. In the case of ordered SQUID arrays we use the same notation 
for the maximum total current, i.e., 
$i_{max}=max\left\{ \frac{1}{N_x N_y} \sum_{n,m} i_{n,m} (t) \right\}$, 
where $n=1,...,N_x$ and $m=1,...,N_y$. For disordered arrays, the brackets
$< ... >$ indicate averaging of the maximum total current of the array over 
the number of differnt realizations, $n_r$.
In order to trace the current-frequency curves for ordered ($i_{max} -\Omega$) 
or disordered ($<i_{max}> -\Omega$) arrays,
we start the system with zero initial conditions and start integrating 
with a low (high) frequency until a steady state is reached. Subsequently the 
frequency is increased (decreased) by a small amount and the equations are
again integrated until a steady state is reached, and so on. In each frequency
step (except the first one, where we use zeros) we use as initial condition the 
steady state solution obtained in the previous step.
\begin{figure}[!h]
\includegraphics[angle=-0, width=0.80\linewidth]{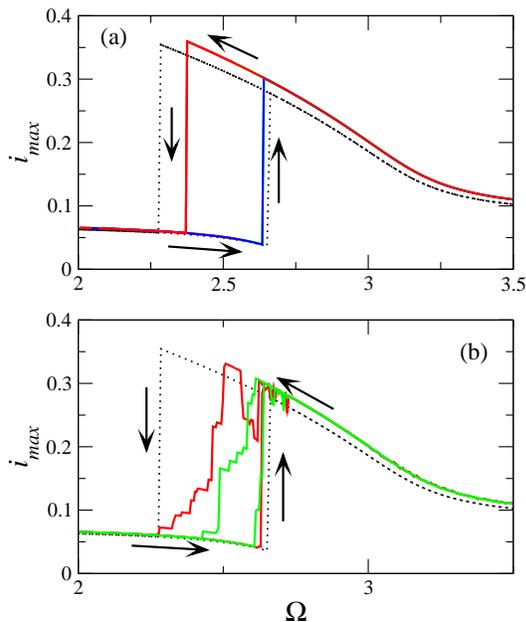}
\caption{(color online)
Maximum total current amplitude $i_{max}$ as a function of the driving frequency
$\Omega$ for two-dimensional $N_x \times N_y$ rf SQUID arrays with $\alpha=0.002$, 
$\beta=1.27$, $\phi_{dc} =0$, $\phi_{ac} =0.1$, $N_x=N_y=20$, and
(a) periodic boundary conditions; 
(b) free-end boundary conditions, starting with different initializations.
The black dotted lines indicate the corresponding $i_{max}$ vs. $\Omega$
curves for a single rf SQUID. The maximum total current of the arrays has been
divided by the total number of rf SQUIDs $N_x \times N_y$ to facilitate the
comparison.
}
\end{figure}
The bistability properties of single SQUID oscillators (see figure 2) are also 
seen in the arrays as well. 
We assume a moderate size array with $N_x=N_y=20$, for which the 
coupling between SQUIDs is isotropic and weak, i.e., $\lambda_x =\lambda_y =-0.014$.
Typical current-frequency curves are shown in figure 5, where the maximum of the 
total current, divided by the total number of SQUIDs in the array, is displayed 
as function of the frequency $\Omega$ of an alternating flux (dc flux is set to zero).   
In this figure, the value of $\beta-$parameter has been selected so that hysteretic
effects are rather strong ($\beta_L \simeq 8$). We observe that bistability appears
in a frequency region of significant width. The corresponding curves for a single 
SQUID are also shown for comparison.
In figure 5(a), where periodic boundary conditions have been employed,   
we observe that although the bistability region for the array is narrower than
that for a single SQUID, the maximum current per SQUID is slightly larger
than that for a single SQUID.
In the case of periodic boundary conditions, the size of the array does not affect 
those results; current-frequency curves for larger arrays with $N_x =N_y=40$
and $N_x =N_y=80$ (not shown) are practically identical to these shown in figure 5(a).
In figure 5(b) and the rest of the paper free-end boundary conditions 
[equations (\ref{2.22})] have been used. In this case, the current-frequency
curves are very sensitive to the initial conditions, the frequency step, and 
other parameters. The curves in figure 5(b) shown in different colours (red and green)
correspond to different initializations of the same system.
\begin{figure}[!h]
\includegraphics[angle=-0, width=0.80\linewidth]{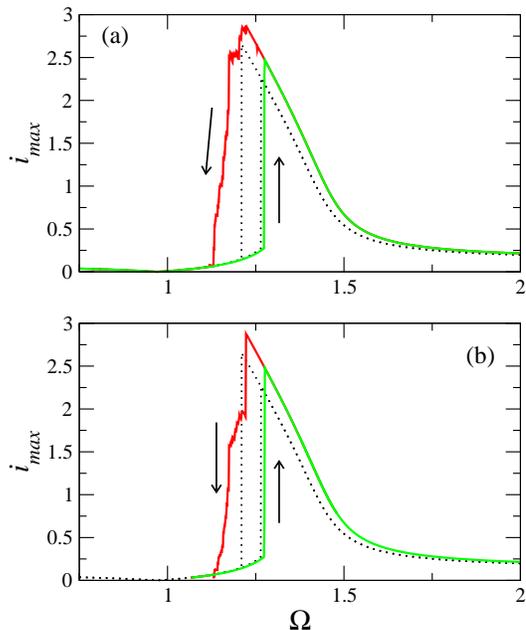}
\caption{(color online)
Maximum total current amplitude $i_{max}$ as a function of the driving frequency
$\Omega$ for two-dimensional $N_x \times N_y$ rf SQUID arrays with $\alpha=0.002$, 
$\beta=0.15$, $\phi_{dc} =0$, $\phi_{ac} =0.02$, and
(a) $N_x=N_y=20$; (b) $N_x=N_y=40$. 
The black dotted lines indicate the corresponding $i_{max}$ vs. $\Omega$
curves for a single rf SQUID. The maximum total current of the arrays has been
divided by the total number of rf SQUIDs $N_x \times N_y$ to facilitate the
comparison. Free-end boundary conditions that account for the termination of the 
structure have been used.
}
\end{figure}

The parts of current-frequency curves that are close to those for the single SQUID
are formed by almost homogeneous states, i.e., states with all SQUIDs in the 
high-current or low-current state. 
Homogeneous states are formed easier in the periodic systems
[figure 5(a)]; only a small part survives in the case of free boundary conditions
[figure 5(b)] that is close to the current-frequency curve of the single SQUID.
In the latter figure we also observe the formation of small steps for which the 
corresponding solutions may be characterized as 'mixed states', that are formed
by a certain number of SQUIDs in the high-current state while all the others are
in the low-current state.
In figure 6, the corresponding curves for SQUIDs with $\beta_L \simeq 1$
($\beta =0.15$) are shown. A comparison with the corresponding curves for a single 
SQUID (shown as dotted black lines) indicates that the frequency region where 
bistability appears are nearly of the same width.
\begin{figure}[!t]
\includegraphics[angle=0, width=0.80\linewidth]{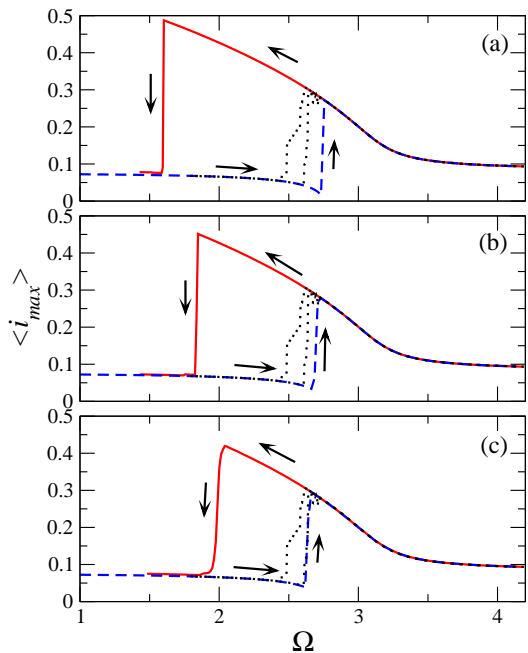}
\caption{(color online)
Maximum current amplitude averaged over $N_R =30$ realizations od disorder, $<i_{max}>$,
as a function of the driving frequency $\Omega$ for a SQUID metamaterial with
$\alpha=0.002$, $\beta=1.27$, $\phi_{ac} =0.1$, $\phi_{dc} =0$, and
(a) $\beta=1.27 ~\pm 0.01$; (b) $\beta=1.27 ~\pm 0.05$; (c) $\beta=1.27 ~\pm 0.1$.
The corresponding curves for the same SQUID metamaterial without disorder are
shown in black-dotted lines.
}
\end{figure}

The assumption of SQUID-based metamaterials comprising identical elements
is certainly not realistic. Therefore, we consider disordered SQUID arrays in
which the parameter $\beta$ varies randomly within a particular range of values 
around a mean value. The SQUID parameter $\beta$, that depends on the critical 
current of the Josephson junctions, determines also the resonance frequency of 
individual SQUIDs. We have calculated the maximum current-frequency curves of
the same SQUID metamaterials as before, taking into account the distribution
of the natural frequencies of individual SQUIDs.
For obtaining reliable results, we have taken statistical averages over 
many realizations, $n_R$, of disorder. Remarkably, the calculations reveal that
weak disorder strongly favours bistability, as it is observed in figures 7 and 8
for mean values of $\beta=1.27$ and $0.15$, respectively. In these figures, 
as we go from (a) to (c) $\beta$ fluctuates by $\pm 0.01$, $\pm 0.05$, and 
$\pm 0.1$, so that the relative disorder strength is much larger in figure 8.  
In all cases shown, either exhibiting weak or strong disorder, the stability
of the nearly homogeneous states with high current increases considerably with
respect to that for the corresponding ordered SQUID metamaterials.   
This is actually reflected in the widening of the bistability regions of the 
current frequency curves.   
We should also note that the bistability region gradually shrinks with increasing 
strength of disorder. That shrinking occurs more rapidly for $\beta_L \simeq 1$
(figure 8) since the relative $\beta$ variation is larger in this case.
\begin{figure}[!t]
\includegraphics[angle=0, width=0.80\linewidth]{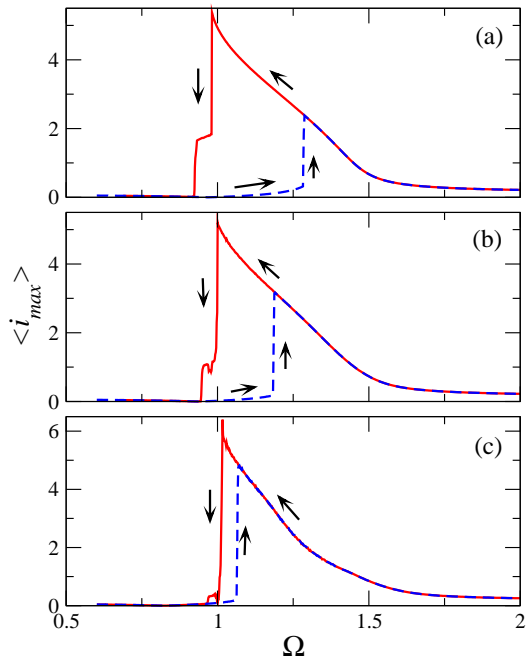}
\caption{(color online)
Maximum current amplitude averaged over $N_R =30$ realizations od disorder, $<i_{max}>$,
as a function of the driving frequency $\Omega$ for a SQUID metamaterial with
$\alpha=0.002$, $\beta=0.15$, $\phi_{ac} =0.02$, $\phi_{dc} =0$, and
(a) $\beta=1.27 ~\pm 0.01$; (b) $\beta=1.27 ~\pm 0.05$; (c) $\beta=1.27 ~\pm 0.1$.
The corresponding curves for the same SQUID metamaterial without disorder are
shown in black-dotted lines.
}
\end{figure}

\section{Synchronization and Multiresponse}
In order to ensure that the maximum current-frequency curves presented in figures
7 and 8 correspond to homogeneous states, we define and calculate a (Kuramoto-type)
synchronization parameter, as
\begin{equation}
\label{4.01}
  \Psi = \left< \frac{1}{N_x \, N_y} \sum_{n,m} e^{2 \pi i \phi_{n,m}} \right>_{\tau,n_R} ,
\end{equation} 
where the brackets denote averaging both in time (i.e., in one oscillation period)
and the number of realizations of disorder $n_R$. The absolute value of $\Psi$ 
quantifies the degree of synchronization; $|\Psi|$ may vary between $0$ and $1$,
corresponding to completely asynchronous and synchronized states, respectively. 
The calculated values for (parts of) the maximum current-frequency curves in figure 7
are shown in figure 9 for $\Omega$ varying in both directions. The bistability 
regions are shown in this figure as green-dotted vertical lines, for reference.
For very weak disorder [figure 9(a)], 
$|\Psi|$ remains close to unity for most of the frequency
interval shown. However, there is a narrow region at low frequencies where 
synchronization, and therefore complete homogeneity breaks down,
when the high current solution loses its stability.
In that case all the SQUIDs change their state towards a lower maximum current 
state.
During the proccess, the phase differences of the SQUID fluxes and the driving 
field lock in randomly selected values, resulting in a low current and only
partially synchronized state. 
However, the two low current states that can be distinguish in this frequency region,
i.e., the synchronized one obtained with increasing frequency and the partially 
synchronized one obtained with decreasing frequency, provide almost the same 
maximum current.
With increasing the strength of the disorder [figure 9(b)] the bistability 
region shinks while the same effect as in figure 9(a) is observed in a wider 
frequency interval. Moreover, the high current states are not completely synchronized
in the bistability region, since $|\Psi|$ is slightly less than unity.
The effect can be seen more clearly by further increasing the strength of the 
disorder as in figure 9(c), where  $|\Psi|$ is clearly less than unity in the 
bistability region indicating partial synchronization.
\begin{figure}[!t]
\includegraphics[angle=0, width=0.80\linewidth]{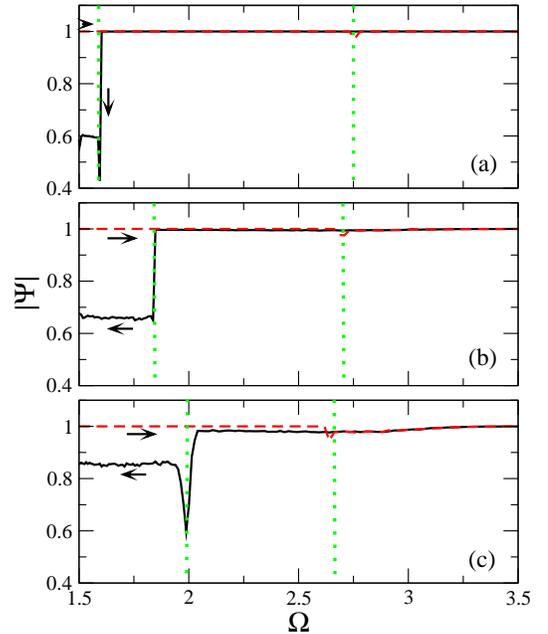}
\caption{(color online)
The magnitude of the synchronization parameter $|\Psi|$ as a function of the driving
frequency $\Omega$ in the bistability region, for a $N_x \times N_y =20 \times 20$
SQUID metamaterial with $\alpha=0.002$, $\beta=1.27 \pm 0.01$, $\phi_{ac} =0.1$, 
$\phi_{dc} =0$, $N_R =30$ realizations of disorder, and 
(a) $\beta=1.27 \pm 0.01$; (b) $\beta=1.27 \pm 0.05$; $\beta=1.27 \pm 0.1$.
The arrows indicate the direction of frequency variation while the green dotted lines
the corresponding bistability intervals.
}
\end{figure}

The magnetic response of the SQUID metamaterial at a particular state can be calculated
in terms of the magnetization along the lines given in references 
\cite{Lazarides2007,Jung2013}. Assuming a tetragonal unit cell with $d_x=d_y=d$ and a
squared SQUID area of side $\alpha$, the magnetization is 
\begin{equation}
\label{4.02}
  M = \frac{\alpha^2 <I>}{d^2 D},
\end{equation} 
where $<I> =I_c \, <i> \equiv I_c \frac{1}{N_x N_y} \sum_{n,m} <i_{n,m}>_\tau$ is the 
spatially and temporally averaged current,
and $D$ is a length related to the cavity where the metamaterial is placed
\cite{Jung2013}. Using fundamental relations of electromagnetism we write the 
relative magnetic permeability as 
\begin{equation}
\label{4.03}
  \mu_r = 1 +\frac{M}{H} , 
\end{equation} 
where $H$ is the intensity of a spatially uniform magnetic field applied 
perpendicularly to the SQUID metamaterial plane.
The latter is related to the external flux to the SQUIDs as
\begin{equation}
\label{4.04}
  H = \frac{\Phi_0}{\mu_0 \alpha^2} <\phi_{ext}> ,
\end{equation} 
where $\mu_0$ is the magnetic permeability of the vacuum, and the brackets denote
temporal averaging. Combining equations (\ref{4.02})-(\ref{4.04}), we get
\begin{figure}[!t]
\includegraphics[angle=0, width=0.80\linewidth]{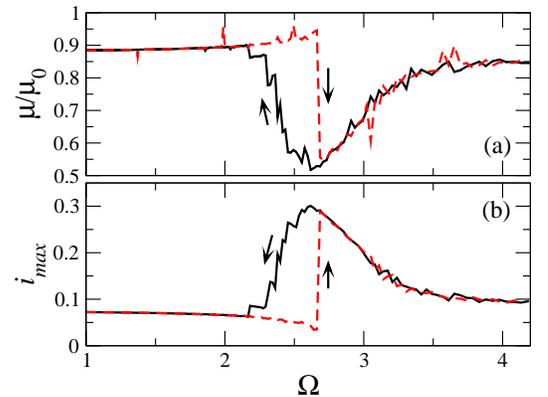}
\caption{(color online)
(a) Relative magnetic permeability $\mu_r =\mu/\mu_0$ for the low and high maximum
current states as a function of the driving frequency $\Omega$, for 
$N_x = N_y =20$, $\alpha=0.002$, $\beta=1.27$, $\phi_{ac} =0.1$, and $\phi_{dc} =0$.
Multiple-valued magnetic response is observed in the bistability region. 
(b) The corresponding maximum current-frequency curves.  
The arrows indicate the direction of frequency variation.
}
\end{figure}
\begin{equation}
\label{4.05}
  \mu_r = 1 +\kappa \frac{<i>}{<\phi_{ext}>} , 
\end{equation} 
where $\kappa =\frac{\mu_0 \alpha I_c}{\Phi_0} \frac{\alpha^3}{d^2 D}$. 
For a rough estimation of the constant $\kappa$ assume that $L \sim \mu_0 \alpha$,
where $L$ is the SQUID inductance, and that $D \simeq d$. Then, we have that
$\kappa \sim \beta \left( \frac{\alpha}{d} \right)^3$. Using $\alpha =d/2$ and 
$\beta =1.27$ we get $\kappa \simeq 0.16$.
We may then use the numerically calculated values of $i_{n,m}$ and $\phi_{ext}$
into equation (\ref{4.05}) to obtain $\mu_r$. 
Simultaneously stable SQUID metamaterial states respond differently to the 
external field and therefore exhibit different $\mu_r$. This can be seen clearly
in figure 10, where the relative permeability $\mu_r$ has been calculated 
from equation (\ref{4.05}). The SQUID metamaterial for the parameters used in
figure 10 is diamagnetic for all frequencies; however, the diamagnetic response 
is stronger for the high current states in the bistability region.
For a weaker driving field that provides a flux amplitude of $10^{-3}$
the metamaterial is in the linear limit, as can be infered by inspection of the 
current-frequency curve shown in figure 11(b). The corresponding $\mu_r$ as a 
function of the driving frequency $\Omega$ is again diamagnetic everywhere
except close to the resonance, where strong variation of $\mu_r$ occurs. 
For frequencies below (but very close to) the resonance at $\Omega \sim 3$
the metamaterial becomes strongly paramagnetic. To the contrary, for frequencies
above (but very close to) the resonance the metamaterial becomes extremely
diamagnetic, exhibiting negative $\mu_r$ within a narrow frequency region.
Note that in this case negative $\mu_r$ would have also been obtained with 
a much smaller coefficient $\kappa$.
\begin{figure}[!t]
\includegraphics[angle=0, width=0.80\linewidth]{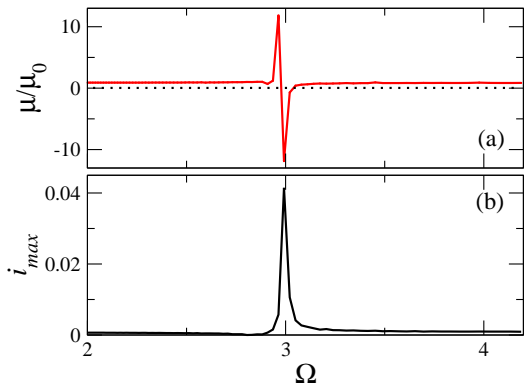}
\caption{(color online)
(a) Relative magnetic permeability $\mu_r =\mu/\mu_0$ for the low and high maximum
current states as a function of the driving frequency $\Omega$, for 
$N_x = N_y =20$, $\alpha=0.002$, $\beta=1.27$, $\phi_{ac} =0.001$, and $\phi_{dc} =0$.
Negative $\mu_r$ is observed in a narrow frequency band just above the resonance
frequency.
(b) The corresponding maximum current-frequency curves. 
}
\end{figure}
\begin{figure}[!t]
\includegraphics[angle=0, width=0.80\linewidth]{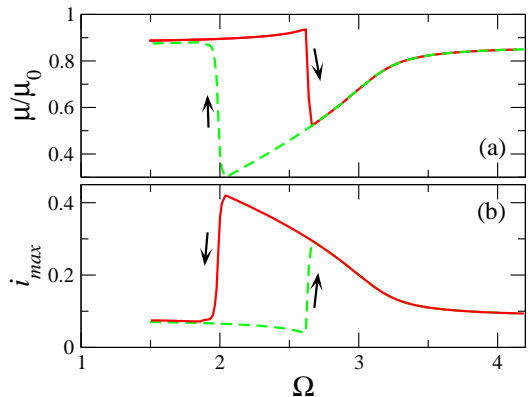}
\caption{(color online)
(a) Relative magnetic permeability $\mu_r =\mu/\mu_0$ for the low and high maximum
current states as a function of the driving frequency $\Omega$, for a disordered
SQUID metamaterial, with 
$N_x = N_y =20$, $\alpha=0.002$, $\beta=1.27 \pm 0.1$, $\phi_{ac} =0.1$, 
and $\phi_{dc} =0$.
Multiple-valued magnetic response is observed in the bistability region. 
(b) The corresponding maximum current-frequency curves. 
}
\end{figure}


For a disordered SQUID metamaterial, that is relatively strongly driven 
($\phi_{ac}=0.1$),
the relative permeability $\mu_r$ changes only slightly. The most important effect 
observed 
in this case is the enlargement of the bistability interval (figure 12).
For weakly driven, disordered SQUID metamaterial, increasing disorder results in
decreasing of the magnitude of $\mu_r$ in the frequency region around resonance.
This effect is illustrated in figures 13 and 14, obtained for two different values
of disorder; $\beta$ fluctuates around its mean value by $\pm 0.01$ and $\pm 0.1$,
respectively. In the frequency region around resonance, in particular, we observe
in figure 13 that the dip corresponding to negative $\mu_r$ becomes shallower than
that for the ordered SQUID metamaterial (figure 11). With further increasing 
disorder (figure 14), the negative $\mu_r$ region disappears. For this particular 
set of parameters, the minimum of $\mu_r$ touches the zero axis.

\begin{figure}[!t]
\includegraphics[angle=0, width=0.80\linewidth]{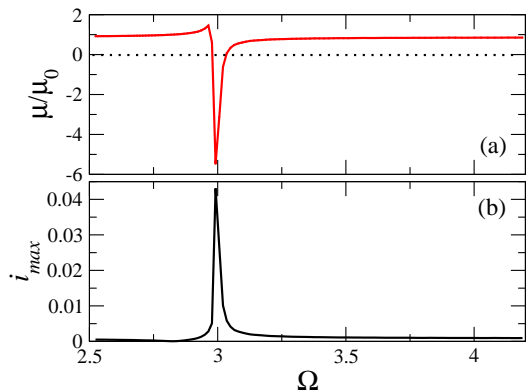}
\caption{(color online)
(a) Relative magnetic permeability $\mu_r =\mu/\mu_0$ for the low and high maximum
current states as a function of the driving frequency $\Omega$, for a disordered
SQUID metamaterial, with 
$N_x = N_y =20$, $\alpha=0.002$, $\beta=1.27 \pm 0.01$, $\phi_{ac} =0.001$, 
and $\phi_{dc} =0$.
Negative $\mu_r$ is observed in a narrow frequency band just above the resonance
frequency.
(b) The corresponding maximum current-frequency curves. 
}
\end{figure}
\begin{figure}[!t]
\includegraphics[angle=0, width=0.80\linewidth]{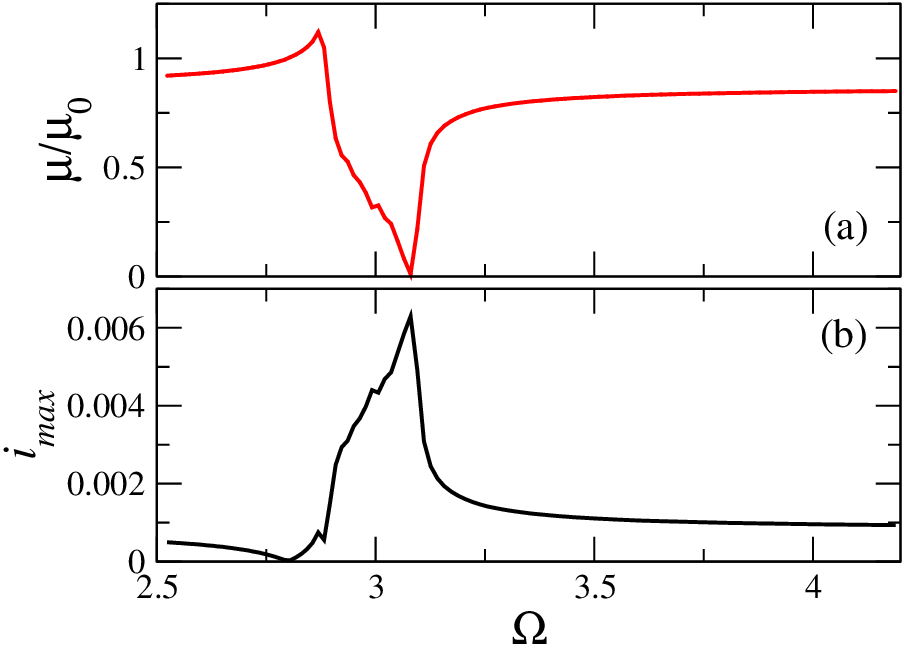}
\caption{(color online)
(a) Relative magnetic permeability $\mu_r =\mu/\mu_0$ for the low and high maximum
current states as a function of the driving frequency $\Omega$, for a disordered
SQUID metamaterial, with 
$N_x = N_y =20$, $\alpha=0.002$, $\beta=1.27 \pm 0.1$, $\phi_{ac} =0.001$, 
and $\phi_{dc} =0$.
Negative $\mu_r$ is not observed because of the relatively stong disorder.
(b) The corresponding maximum current-frequency curves. 
}
\end{figure}


\section{Conclusions.}
We investigated numerically two-dimensional SQUID metamaterials driven by an 
alternating magnetic field. We have calculated current-frequency curves both for 
ordered and disordered metamaterials; in both cases we observed bistability regions
created by almost homogeneous high and low current states. However, we observed 
that the presence of disorder widens significantly the bistability regions.
The effect is rather strong for SQUID metamaterials comprising SQUIDs with either
small or large $\beta_L$ parameter. Remarkably, the homogeneity of high and low 
current states, that is quantified by the parameter $\Psi$, persists also in the 
case of disorder up to a very high degree. These states deviate only slightly from
practically complete homogeneity only in the case of strong disorder, as can be
seen from calculations of the amplitude of $\Psi$. This important result indicates 
that random variation of the SQUID parameters does not destroy bistability,
that is crusial for applications that require bistable switching properties, 
but instead stabilizes the system against modulational or other instabilities.

This result is related to past work on disordered networks of nonlinear oscillators
where it was concluded that moderate disorder may enhance synchronization and
stabilize the system against chaos \cite{Braiman1995a,Braiman1995b}.
In the present context, synchronization of individual SQUIDs in the high or low
current states results in high or low maximum total current for the metamaterial.
This requires that (almost) all the SQUIDs are in phase. It could be natural to 
assume that the more nearly identical the elements, the better the synchronization
will be. However, even in the ideal case of identical elements, the earlier 
assumption may not be true and the in phase state may be dynamically unstable.
Then, synchronization is reduced and the SQUID metamaterial cannot remain in the
high current state that is more sensitive to instability. This can be clearly 
observed in figures 5(b) and 6(b), where in most of the bistability region the
metamaterial relaxes to partially synchronized states that provide significantly
lower maximum total current. This type of disorder-assisted self-organization
may also occur by introducing local disorder in an array of otherwise identical
oscillators, i.e., in the form of impurities \cite{Gavrielides1998a,Gavrielides1998b}.
In this case, the impurities trigger a self-organizing process that brings
the system to complete synchronization and suppression of chaotic behaviour.
   
Having calculated numerically the response of the metamaterial to an alternating 
field with given frequency, we have calculated the magnetic permeability of the 
metamaterial for several illustrating cases both with and without disorder.
While the expression for calculating the magnetic permeability is rather simple,
there is some uncertainty about the value of the factor $\kappa$. However, for 
a reasonable value of $\kappa$ we observe that the magnetic permeability can take
negative values in a narrow frequency region above resonance for weakly driven 
SQUID metamaterials. In this case, increasing disorder results in weakening the
negative response of the metamaterial; thus, for relatively strong disorder the 
response is not sufficient to make the magnetic permeability negative. 
For SQUID metamaterials exhibiting bistability, different magnetic permeabilities 
can be reached under the same conditions depending on their state.

\section*{Acknowledgements.
}
This research was partially supported by the THALES Project MACOMSYS,
co-financed by the European Union (European Social Fund – ESF)
and Greek national funds through the Operational Program 
"Education and Lifelong Learning" of the National Strategic Reference Framework 
(NSRF) - Research Funding Program: THALES. 
Investing in knowledge society through the European Social Fund.

\bibliography{MyBibTex_Library_15Mar2013.bib}
\bibliographystyle{unsrt}
\end{document}